\documentstyle[]{europhys}

\def\And{{\rm and\ }}

\newif\ifboo \boofalse

\begin{document}

\euro{XX}{X}{1-$\infty$}{2000}
\Date{19 May 2000}
\shorttitle{E. M. Chudnovsky and D. A. Garanin COMMENT ON LEUENBERGER AND LOSS}
\title{ 
 Crystal Field $-AS_{z}^{2}$ Does Not Produce
One-Phonon Transitions With ${\Delta}S_{z}={\pm}2$ [Comment on EPL 46, 692 (1999) by Leuenberger and Loss] 
}      

\author{
E. M. Chudnovsky\inst{1} 
\footnote{chudnov@lehman.cuny.edu},
\And D. A. Garanin\inst{1}
} 
\institute{
     \inst{1} Physics Department, CUNY Lehman College\\ Bedford Park Boulevard West, Bronx, NY 10468-1589
}
%
%
\rec{}{}
%
\pacs{
\Pacs{75}{45.+j}{Macroscopic quantum phenomena in magnetic systems}
\Pacs{75}{50.Tt}{Fine-particle systems}
}
\maketitle

\vspace{-1cm}

Recently Leuenbeger and Loss suggested a theory of phonon-assisted relaxation in a molecular nanomagnet Mn-12 that ``contrary to previous results is in reasonably good agreement not only with the relaxation data but also with all experimental parameter values known so far'' \cite{LL-EPL,LL-PRB}. The purpose of this Comment is to show that the model of Leuenberger and Loss and its comparison with experiment are premised upon a principal error.

The theory of Ref.\ \cite{LL-EPL} follows the footsteps of our earlier work \cite{GC} that describes phonon-assisted magnetic relaxation in Mn-12 in terms of the master equation for the density matrix. The dominant term in the crystal field of Mn-12 is 
$-AS_{z}^{2}$. Performing rotation of the anisotropy axis due to the elastic deformation ${\bf u}$, one obtains the following magnetoelastic coupling \cite{GC}
\begin{equation}
A({\omega}_{xz}\{S_{x},S_{z}\}+{\omega}_{yz}\{S_{y},S_{z}\}=
 (A/2) [({\omega}_{xz}-i{\omega}_{yz})\{S_{+},S_{z}\}+
({\omega}_{xz}+i{\omega}_{yz})\{S_{-},S_{z}\}]\;\;,
\end{equation}
where ${\omega}_{{\alpha}{\beta}}=\frac{1}{2}
({\partial}_{\beta}u_{\alpha}-{\partial}_{\alpha}u_{\beta})$. Operators $S_{\pm}$ change the $S_{z}$ projection of spin by ${\pm}1$, while ${\bf u}$ is linear on the operators of creation and annihilation of phonons. Consequently, Eq.\ (1), when quantized, describes emission and absorption of one phonon accompanied by
${\Delta}S_{z}={\pm}1$. 

According to Leuenberger and Loss \cite{LL-EPL,LL-PRB}, the crystal field $-AS_{z}^{2}$ also produces the magnetoelastic coupling of the form 
\begin{equation}
A({\epsilon}_{xx}-{\epsilon}_{yy})(S_{x}^{2}-S_{y}^{2})= 
(A/2)({\epsilon}_{xx}-{\epsilon}_{yy})(S_{+}^{2}+S_{-}^{2})\;\;.
\end{equation}
where ${\epsilon}_{{\alpha}{\beta}}$ is the strain tensor. This coupling generates ``second-order'' spin-phonon transitions with ${\Delta}S_{z}={\pm}2$, which ``lead to a much faster relaxation of the spin system'' than ``first-order'' transitions with ${\Delta}S_{z}={\pm}1$. 

The error stems from the use by Leuenbeger and Loss \cite{LL-EPL,LL-PRB} of the linear formula for the strain rensor,
\begin{equation} 
{\epsilon}_{{\alpha}{\beta}}=\frac{1}{2}\left(\frac{{\partial}u_{\alpha}}
{{\partial}x_{\beta}}+\frac{{\partial}u_{\beta}}
{{\partial}x_{\alpha}}\right)\;\;,
\end{equation}
instead of the exact expression \cite{Landau},
\begin{equation} 
{\epsilon}_{{\alpha}{\beta}}=\frac{1}{2}\left(\frac{{\partial}u_{\alpha}}
{{\partial}x_{\beta}}+\frac{{\partial}u_{\beta}}{{\partial}x_{\alpha}}+
\frac{{\partial}u_{\gamma}}{{\partial}x_{\alpha}}
\frac{{\partial}u_{\gamma}}{{\partial}x_{\beta}}\right)\;\;. 
\end{equation}
To obtain Eq.\ (2) Leuenberger and Loss computed the rotation matrix $\hat{R}$ and the corresponding deformation, 
${\bf u}=({\hat{R}}_{z}{\hat{R}}_{y}{\hat{R}}_{x}-1){\bf x}$, up to the {\it second order} in infinitesimal rotation 
$\delta\phi_\alpha$. They obtained \cite{LL-EPL,LL-PRB} 
\begin{eqnarray} 
u_{x} & = & {\delta}{\phi}_{y}z-{\delta}{\phi}_{z}y-
(1/2)({\delta}{\phi}_{y}^{2}+{\delta}{\phi}_{z}^{2})x  
\nonumber \\
u_{y} & = & {\delta}{\phi}_{z}x-{\delta}{\phi}_{x}z-
(1/2)({\delta}{\phi}_{x}^{2}+{\delta}{\phi}_{z}^{2})y  
\nonumber \\
u_{z} & = & {\delta}{\phi}_{x}y-{\delta}{\phi}_{y}x-
(1/2)({\delta}{\phi}_{x}^{2}+{\delta}{\phi}_{y}^{2})z  
\end{eqnarray}
The incorrect Eq.\ (3) then gives (incorrectly)
\begin{equation}
{\epsilon}_{xx} =-({\delta}{\phi}_{y}^{2}+{\delta}{\phi}_{z}^{2})/2,
\qquad 
{\epsilon}_{yy} = -({\delta}{\phi}_{x}^{2}+{\delta}{\phi}_{z}^{2})/2,
\qquad 
{\epsilon}_{zz} = -({\delta}{\phi}_{x}^{2}+{\delta}{\phi}_{y}^{2})/2,
\end{equation}  
and one gets ${\delta}{\phi}_{x}^{2}={\epsilon}_{xx}-{\epsilon}_{yy}-
{\epsilon}_{zz}$ and cyclic permutations for ${\delta}{\phi}_{y}^{2}$ and ${\delta}{\phi}_{z}^{2}$. Substituting this into $\hat{R}$ and inserting the rotated spin vector $R_{x}R_{y}{\bf S}$ into 
$-AS_{z}^{2}$, Leuenberger and Loss obtained Eq.\ (2). One should notice, however, that the substitution of Eq.\ (5) into the correct expression for the strain tensor, Eq.\ (4), yields
\begin{equation} 
{\epsilon}_{xx}={\epsilon}_{yy}={\epsilon}_{zz}=0\;\;,
\end{equation}
in accordance with the fact that rotation does not change the volume, ${\delta}V=V\sum \epsilon_{\alpha\alpha}=0$. 

In fact, to see that Eq.\ (2) cannot be right no calculation is needed. Indeed, the operator $-AS_{z}^{2}$ conserves $S_{z}$. In order to change the magnetic quantum number of Mn-12 molecule by 2, one would have to assign to phonons spin 2. Phonons, however, being described by a vector field ${\bf u}$, cannot possess spin other than 1 \cite{Messiah}. One-phonon processes accompanied by ${\Delta}S_{z}={\pm}2$ can only be generated by terms in the crystal field which do not conserve $S_{z}$. For Mn-12 these are tunneling terms which are orders of magnitude smaller than the uniaxial crystal field. Leuenberger and Loss (see Appendix D of Ref.\ \cite{LL-PRB}) find support of their incorrect statement on p.\ 563 of Abragam and Bleaney \cite{AB}. However, Abragam and Bleaney, when talking about ``first-'' and ``second-order'' transitions, mean two-phonon Raman processes, which, of course, are not prohibited by the above argument. These two-phonon processes die out at low temperature. Their rate is inversely proportional to the tenth power of the sound velocity, as compared to the fifth power for the one-phonon processes. Leuenberger and Loss compare their theory with experiment based upon the extraction of the sound velocity from the measured relaxation rate,
using a wrong type of the phonon process. Their claim of agreement with experiment is, therefore, completely invalidated by their incorrect model for the spin-phonon coupling. 

This work has been supported by the NSF Grant No. DMR-9978882.



\begin{thebibliography}{99}
\bibitem{LL-EPL} M. N. Leuenberger and D. Loss, Europhys. Lett. {\bf 46}, 692 (1999).
\bibitem{LL-PRB} M. N. Leuenberger and D. Loss, Phys. Rev. {\bf B61},
1286 (2000).
\bibitem{GC} D. A. Garanin and E. M. Chudnovsky, Phys. Rev. {\bf B56},
11102 (1997).
\bibitem{Landau} L. D. Landau and E. M. Lifshits, Theory of Elasticity
(Pergammon, New York, 1970).
\bibitem{Messiah} A. Messiah, Quantum Mechanics (de Gruyter, New York, 1991).
\bibitem{AB} A. Abragam and A. Bleaney, Electron Paramagnetic Resonance of Transition Ions (Clarendon Press, Oxford, 1970).
\end{thebibliography}
\end{document}